\documentclass[aps,prb,preprint,showpacs,preprintnumbers,amsmath,amssymb]{revtex4}
\usepackage{amsmath}
\usepackage{amssymb}
\usepackage{amsfonts}
\usepackage{color,array}
\usepackage{dsfont}
\usepackage{slashed}
\usepackage{graphicx}
\usepackage{graphics}
\usepackage{epsfig}
\usepackage{bbm,bm}
\usepackage{psfrag}
\usepackage{ulem}
\usepackage{url}

\begin{document}
\setcounter{footnote}{0}

\title{Critical Exponents in Two Dimensions and Pseudo-$\epsilon$ Expansion}

\author{M. A. Nikitina}
\author{A. I. Sokolov}
\email{ais2002@mail.ru}
\affiliation{Department of Quantum Mechanics,
Saint Petersburg State University,
Ulyanovskaya 1, Petergof,
Saint Petersburg, 198504
Russia}
\date{\today}

\begin{abstract}
The critical behavior of two-dimensional $n$-vector $\lambda\phi^4$ field model
is studied within the framework of pseudo-$\epsilon$ expansion approach.
Pseudo-$\epsilon$ expansions for Wilson fixed point location $g^*$ and critical
exponents originating from five-loop 2D renormalization group series are derived.
Numerical estimates obtained within Pad\'e and Pad\'e-Borel resummation procedures
as well as by direct summation are presented for $n = 1$, $n = 0$ and $n = -1$,
i. e. for the models which are exactly solvable.
The pseudo-$\epsilon$ expansions for $g^*$, critical exponents $\gamma$ and $\nu$
have small lower-order coefficients and slow increasing higher-order ones. As a
result, direct summation of these series with optimal cut off provides numerical
estimates that are no worse than those given by the resummation approaches
mentioned. This enables one to consider the pseudo-$\epsilon$ expansion technique
itself as some specific resummation method.
\end{abstract}


\maketitle

\section{Introduction}

Pseudo-$\epsilon$ expansion is known to be rather effective when used to
estimate numerical values of universal quantities characterizing critical
behavior of three-dimensional systems \cite{GZJ1980,GZJ1998,FHY2000,HID2004}.
Moreover, even in two dimensions, where original renormalization group (RG)
series are shorter and more strongly divergent, pseudo-$\epsilon$ expansion
technique is able to give good or satisfactory results \cite{GZJ1980,COPS2004,
S2005,S2013}. To obtain numerical estimates from pseudo-$\epsilon$ expansions 
one applies a resummation technique since corresponding series have growing
higher-order coefficients, i. e. look divergent. In contrast to RG expansions in
fixed and $4-\epsilon$ dimensions, pseudo-$\epsilon$ expansions do not need in
advanced resummation procedures based on Borel transformation. As a rule, use of
simple Pad\'e approximants turns out to be sufficient to obtain proper numerical
estimates \cite{FHY2000,COPS2004,S2005,S2013}.

In this paper, we study the critical behavior of two-dimensional $O(n)$-symmetric
systems within the frame of pseudo-$\epsilon$ expansion technique. The series for
the Wilson fixed point location $g^*$ and critical exponents originating from the
five-loop RG expansions will be derived for arbitrary order parameter
dimensionality $n$. The pseudo-$\epsilon$ expansions obtained will be analysed in
detail for $n = 1$, $n = 0$ and $n = -1$, i. e. for the models with exactly known 
critical exponents \cite{N1982,N1984,FQS1984}. First of them (the Ising model) 
describes phase transitions in numerous physical systems including uniaxial 
ferromagnets and liquid mixtures while the second corresponds to a long polymer  
in solution\cite{PdG1972} (self-avoiding walks). Three models mentioned may be 
considered as testbeds for clarification of the numerical effectiveness of various 
approximation schemes including RG perturbation theory and the method of 
pseudo-$\epsilon$ expansion. Numerical estimates for critical exponents will be 
extracted from the pseudo-$\epsilon$ expansions by means of Pad\'e and Pad\'e-Borel 
resummation techniques as well as by direct summation. The latter approach will 
be applied under the assumption that the best numerical results may be obtained by 
means of cutting divergent pseudo-$\epsilon$ expansions off by smallest terms, 
i. e. applying the procedure valid for asymptotic series.

\section{Pseudo-$\epsilon$ expansions for general $n$}

The critical behavior of two-dimensional systems with $O(n)$-symmetric vector
order parameters is described by Euclidean field theory with the Hamiltonian:
\begin{equation}
\label{model}
H =
\int d^2x \Biggl[{1 \over 2}( m_0^2 \varphi_{\alpha}^2
 + (\nabla \varphi_{\alpha})^2)
+ {\lambda \over 24} (\varphi_{\alpha}^2)^2 \Biggr] ,
\end{equation}
where $\varphi_{\alpha}$ is a real $n$-vector field, bare mass squared $m_0^2$
being proportional to $T - T_c^{(0)}$, $T_c^{(0)}$ -- phase transition temperature
in the absence of order parameter fluctuations. The $\beta$-function and the
critical exponents for the model (1) have been calculated within the massive
theory \cite{OS2000,COPS2004}, with the Green function, the four-point vertex
and the $\phi^2$ insertion normalized in a conventional way:
\begin{eqnarray}
\label{norm}
G_R^{-1} (0, m, g_4) = m^2 , \qquad \quad
{{\partial G_R^{-1} (p, m, g_4)} \over {\partial p^2}}
\bigg\arrowvert_{p^2 = 0} = 1 , \\
\nonumber
\Gamma_R (0, 0, 0, m, g) = m^2 g_4, \qquad \quad
\Gamma_R^{1,2} (0, 0, m, g_4) = 1.
\end{eqnarray}
Starting from the five-loop RG expansion for $\beta$-function \cite{OS2000},
we replace the linear term in this expansion with $\tau g$, calculate the
Wilson fixed point coordinate $g^*$ as series in $\tau$, and arrive to the
following expression:
\begin{eqnarray}
\label{g-tau}
g^* &=& \tau + {\tau^2 \over (n + 8)^2} \biggl( 10.33501055~n + 47.67505273 \biggr)
\nonumber \\
&+& {\tau^3 \over (n + 8)^4} \biggl(- 5.00027593~n^3 + 24.4708201~n^2 + 253.297221~n
+ 350.808487 \biggr) \ \
\nonumber \\
&+& {\tau^4 \over (n + 8)^6} \biggl( 0.088842906~n^5 - 77.270445~n^4
+ 45.052398~n^3 + 3408.2839~n^2
\nonumber \\
&+& 14721.151~n + 27649.346 \biggr) - {\tau^5 \over (n + 8)^8} \biggl(- 0.00407946~n^7
- 0.305739~n^6
\nonumber \\
&+& 1464.58~n^5 + 11521.4~n^4 + 98803.3~n^3 + 794945~n^2 + 3.14662~10^6~n + 4.73412~10^6 \biggr). \ \
\end{eqnarray}
Substituting this expansion into five-loop RG series for critical exponents $\gamma$
and $\eta$ \cite{OS2000, COPS2004} we obtain:

\begin{eqnarray}
\label{gamma-tau}
\gamma^{-1} &=& 1 - {\tau~ (n + 2) \over (n + 8)}
- {\tau^2 \over (n + 8)^3} \biggl( 6.95938160~n^2 + 34.58878428~n + 41.34004218 \biggr)
\nonumber \\
&+& {\tau^3 \over (n + 8)^5} \biggl(0.338391156~n^4 - 53.7045862~n^3 - 181.874852~n^2
+ 471.838217~n
\nonumber \\
&+& 1236.12490 \biggr) - {\tau^4 \over (n + 8)^7} \biggl(- 0.23015013~n^6 + 21.848143~n^5
+ 1537.3578~n^4
\nonumber \\
&+& 12405.258~n^3 + 41577.259~n^2 + 75410.316~n + 59869.804 \biggr) \ \
\nonumber \\
&+& {\tau^5 \over (n + 8)^9} \biggl(0.115623~n^8 + 17.8566~n^7 + 83.1552~n^6 + 14850.5~n^5
- 84964.7~n^4
\nonumber \\
&+& 318099~n^3 + 3.76620~10^6~n^2 + 1.08837~10^7~n + 1.01285~10^7 \biggr). \ \
\end{eqnarray}

\begin{eqnarray}
\label{eta-tau}
\eta &=& {\tau^2 \over (n + 8)^2}~0.9170859698 \biggl(n + 2 \biggr)
\nonumber \\
&+& {\tau^3 \over (n + 8)^4} \biggl(- 0.0546089776~n^3 + 17.9732248~n^2 + 120.114155~n
+ 167.898539 \biggr) \ \
\nonumber \\
&+& {\tau^4 \over (n + 8)^6} \biggl(- 0.092684458~n^5 - 8.2910597~n^4 + 174.43187~n^3
+ 2120.0408~n^2
\nonumber \\
&+& 7034.6638~n + 7114.3103 \biggr) + {\tau^5 \over (n + 8)^8} \biggl(- 0.0709196~n^7
- 5.60392~n^6 - 250.874~n^5
\nonumber \\
&+& 1312.68~n^4 + 36126.0~n^3 + 201476~n^2 + 470848~n + 396119 \biggr). \ \
\end{eqnarray}
Pseudo-$\epsilon$ expansions for other critical exponents can be deduced from (4), (5)
using well-known scaling relations. The series for the correlation length exponent
$\nu$, for example, results from the formula
\begin{equation}
\label{scaling}
\gamma = \nu (2 - \eta).
\end{equation}

It is worthy to note that in two dimensions only models with $-2 < n < 2$ undergo  
transitions into ordered phase, i. e. into the spatially uniform state with non-zero 
order parameter. From the physical point of view, the series obtained apply to this 
domain of $n$. On the other hand, two-dimensional phase transition models with 
$n \ge 2$ are widely explored\cite{OS2000,BK92,KP93,K95,BC96,CPRV96,PV98,PV02} to 
evaluate numerical power of various lattice and field-theoretical approaches. It 
looks instructive in this context to study $\tau$-series (3)-(5) for $n \ge 2$ as 
well. The first step in this direction has been recently done\cite{NS14c}.

\section{Critical exponents for $n=1$, $n=0$ and $n =-1$}

It is of major interest to analyze numerical results given by the
obtained expansions for the values of $n$ under which the model
(1) is exactly solvable. That is why further we concentrate on the 
cases $n=1$, $n=0$, and $n =-1$. Pseudo-$\epsilon$ expansions for 
critical exponents we'll deal with are as follows:

\begin{center}
\textbf{$n = 1$}
\end{center}
\begin{eqnarray}
\gamma = 1 + \tau/3 + 0.224812357 \tau^{2}
+ 0.087897190 \tau^{3}+0.086443008\tau^{4}-0.0180209 \tau^{5}.
\end{eqnarray}
\begin{eqnarray}
\gamma^{-1} = 1 - \tau/3 - 0.113701246 \tau^2
+ 0.024940678 \tau^3-0.039896059 \tau^4+0.0645210 \tau^5.
\end{eqnarray}
\begin{eqnarray}
\nu = 1/2 + \tau/6 + 0.120897626 \tau^{2}
+ 0.0584361287 \tau^{3} + 0.056891652 \tau^{4} + 0.00379868 \tau^{5}.
\end{eqnarray}
\begin{eqnarray}
\nu^{-1} = 2 - 2\tau/3 - 0.261368281 \tau^{2}
+ 0.0145750797 \tau^{3} - 0.091312521 \tau^{4} + 0.118121 \tau^{5}.
\end{eqnarray}
\begin{eqnarray}
\eta = 0.0339661470 {\tau}^{2}+0.0466287623 {\tau}^{3}
+ 0.030925471 {\tau}^{4}+0.0256843 {\tau}^{5}.
\end{eqnarray}

\begin{center}
\textbf{$n = 0$}
\end{center}
\begin{eqnarray}
\gamma = 1 + \tau/4 + 0.143242270 \tau^{2} + 0.018272597 \tau^{3}
+ 0.035251118 \tau^{4}-0.0634415\tau^{5}.
\end{eqnarray}
\begin{eqnarray}
\gamma^{-1} = 1- \tau/4 - 0.080742270\tau^{2}+0.037723538 \tau^{3}
- 0.028548147 \tau^{4}+0.0754631 \tau^{5}.
\end{eqnarray}
\begin{eqnarray}
\nu = 1/2 + \tau/8 + 0.0787857831 \tau^{2}
+ 0.0211750671 \tau^{3} + 0.028101050 \tau^{4} - 0.0222040 \tau^{5}.
\end{eqnarray}
\begin{eqnarray}
\nu^{-1} = 2 - \tau/2 - 0.190143132 \tau^{2}
+ 0.0416212976 \tau^{3} - 0.071673308 \tau^{4} + 0.136330 \tau^{5}.
\end{eqnarray}
\begin{eqnarray}
\eta = 0.0286589366 {\tau}^{2} + 0.0409908542 {\tau}^{3}
+ 0.027138940 {\tau}^{4} + 0.0236106 {\tau}^{5}.
\end{eqnarray}

\begin{center}
\textbf{$n = -1$}
\end{center}
\begin{eqnarray}
\gamma = 1 + \tau/7 + 0.060380873 \tau^{2}-0.023532210 \tau^{3}
+ 0.012034268 \tau^{4}-0.0638772 \tau^{5}.
\end{eqnarray}
\begin{eqnarray}
\gamma^{-1} = 1- \tau/7 - 0.039972710 \tau^2+0.037868436 \tau^3
- 0.018392201 \tau^{4}+ 0.0649966 \tau^{5}.
\end{eqnarray}
\begin{eqnarray}
\nu = 1/2 + \tau/14 + 0.0348693698 \tau^{2}
- 0.00424514372 \tau^{3} + 0.011608435 \tau^{4} - 0.0268913 \tau^{5}.
\end{eqnarray}
\begin{eqnarray}
\nu^{-1} = 2 - 2\tau/7 - 0.0986611527 \tau^{2}
+ 0.0510003794 \tau^{3} - 0.049264800 \tau^{4} + 0.116842 \tau^{5}.
\end{eqnarray}
\begin{eqnarray}
\eta = 0.0187160402 {\tau}^{2} + 0.0274103364 {\tau}^{3}
+ 0.017144702 {\tau}^{4} + 0.0159901 {\tau}^{5}.
\end{eqnarray}

The expansions for "big" critical exponents $\gamma$, $\nu$ and for their
inverses are seen to possess coefficients which begin to grow from certain
terms indicating that these series are divergent. Moreover, they are not
alternative, i. e. their coefficients have irregular signs. At the same time,
lower-order coefficients in expansions (7)--(10), (12)--(15) and (17)--(20)
decrease, and decrease more rapidly than their counterparts in the original
RG series. This enables one to consider them as suitable for some resummation
and getting proper numerical estimates.

The structure of pseudo-$\epsilon$ expansions for "small" exponent $\eta$ is
quite different. These series have positive coefficients of the same order of
magnitude what makes questionable an applicability of any procedure employed
nowadays for resummation of diverging RG series.

To demonstrate a power of various resummation techniques and to clear up to
what extent they are necessary in the case considered we present below
numerical results given by several relevant procedures. Namely, we evaluate
critical exponents $\gamma$ and $\nu$ for $n=1$, $n=0$ and $n =-1$ by means
of the Pad\'e resummation, by Pad\'e-Borel resummation of the pseudo-$\epsilon$
expansions for exponents themselves and for their inverses, and by direct
summation of the series (7)--(10), (12)--(15) and (17)--(20). Direct summation
is performed under the assumption that one can get the best numerical estimates
cutting off divergent pseudo-$\epsilon$ expansions by smallest terms, i. e.
adopting the procedure true for asymptotic series.

The results thus obtained are collected in Table I. Along with
pseudo-$\epsilon$ expansion estimates the exact values of critical
exponents and the numbers originating from resummed five-loop RG
series \cite{OS2000} are presented for comparison. Numerical
values of the Fisher exponent given by direct summation of series
(11), (16), and (21) are also included to give an idea about the
accuracy of pseudo-$\epsilon$ expansion technique in the case of
small critical exponent.

Before discussing content of Table 1 we present some details concerning the
critical exponent values obtained. In principle, Pad\'e resummation procedure is
known to be rather effective when applied to pseudo-$\epsilon$ expansions for
critical exponents and other universal quantities
\cite{FHY2000, HID2004, COPS2004, S2005}. It demonstrates, as a rule, good
convergence if one works within high enough orders in $\tau$. In two dimensions,
however, the numbers given by Pad\'e resummed expansions may converge to the
values differing considerably from their exact counterparts. Pad\'e triangles
presented below illustrate this situation. The first one (Table II) shows most
favorable situation - exponent $\nu$ at $n = 0$ - when numerical estimates
regularly converge to the true value $\nu = 0.75$. The second example (Table III)
demonstrates that good convergence may not result in quite good numerical
estimate: the asymptotic value $\nu = 0.606$ differs appreciably from the exact
one $\nu = 0.625$ for $n = -1$. At last, from Table IV (the exponent $\gamma$,
$n = 0$) one can see that fair convergence does not guarantee satisfactory
numerical results - the estimates in this Table concentrate near 1.435, i. e.
far from the exact value 1.34375.

Similar situation takes place when we address Pad\'e-Borel resummation technique.
This procedure may result in either good numerical results or unsatisfactory
ones depending on the critical exponent evaluated and on the value of $n$.
Tables V-VII illustrate this statement. Pad\'e-Borel resummation of the
pseudo-$\epsilon$ expansion of the inverse exponent $\nu$ for $n = 0$ gives
quite good numerical estimates (Table V) while estimates of $\nu$ for $n = -1$
and $\gamma$ for $n = 0$ via inverse expansions (Tables VI and VII) "miss" the
exact values. Moreover, Pad\'e-Borel triangles for exponents $\gamma$ and $\nu$
themselves at $n = 1$ and some others turn out to be half-empty because many
Pad\'e approximants are spoilt by "dangerous" (positive axis) poles.

\section{To resum or not to resum?}

Let us return back to Table I. As is seen, numerical estimates provided by Pad\'e
and Pad\'e-Borel resummation techniques may i) be considerably scattered and
ii) differ from the exact values no less than numbers given by direct summation
of pseudo-$\epsilon$ expansions and of corresponding inverse series. On the other
hand, direct summation of these expansions generates an iteration procedure which
rapidly converge to asymptotic values that are as close to the exact ones as those
obtained within various resummation methods. Figures 1-4 where partial sums of
series (7-10) and (12-15) are depicted as functions of $k$, $k$ being the order in 
$\tau$, illustrate the situation. Filled rounds and triangles mark the points of 
optimal cut off, i. e. the order from which the coefficients of pseudo-$\epsilon$ 
expansions start to grow. Figures 1, 2 show the favorable cases when approximate 
values almost coincide with exact ones. Figures 3, 4, to the contrary, show most 
unfavorable regimes when the difference between approximate and exact values turns 
out to reach 0.1. Analogous level of accuracy is observed when small critical 
exponent $\eta$ is estimated. Indeed, the direct summation of the pseudo-$\epsilon$ 
expansion (see Table I) and application of the resummation techniques result in 
numbers grouping around the exact values within the range of order of 0.1.

So, the resummation of pseudo-$\epsilon$ expansions for two dimensional models
practically does not improve numerical estimates of critical exponents. Moreover,
the direct summation leads to approximate values which are as accurate as those
resulting from original five-loop RG series (see Table I). This enables us to
conclude that estimating critical exponents in two dimensions within the
pseudo-$\epsilon$ expansion approach one can use the simplest possible way to
process the series - direct summation with optimal cut off\cite{3D}. 
 
In this sense the pseudo-$\epsilon$ expansion itself may be considered as some
special resummation method. There are two reasons for such a point of view. First,
this approach transforms strongly divergent field-theoretical RG expansions into
power series with much smaller lower-order coefficients and much slower increasing
higher-order ones. Second, the physical value of expansion parameter $\tau$ is
equal to 1, while the Wilson fixed point coordinate $g^*$ playing analogous role
within field-theoretical RG approach is almost two times bigger in two dimensions
($g^* = 1.84 - 1.86$ for $n = 1, 0, -1$ \cite{OS2000}). This difference looks
essential, especially keeping in mind importance of higher-order terms.

\section{Conclusion}

To summarize, we have calculated pseudo-$\epsilon$ expansions for dimensionless
effective coupling constant $g^*$ and critical exponents of 2D Euclidean $n$-vector
field theory up to $\tau^5$ order. Numerical estimates of critical exponents for
models with $n = 1, 0, -1$ exactly solvable at criticality have been found using
Pad\'e and Pad\'e-Borel resummation techniques as well as by direct summation with
optimal cut off. Comparison of the results obtained with each others and with their
exact counterparts has shown that direct summation of pseudo-$\epsilon$ expansions
provides, in general, numerical estimates that are no worse than those given by
resummation approaches mentioned. This implies that the pseudo-$\epsilon$ expansion
approach itself may be thought of as some specific resummation technique.

\section*{ACKNOWLEDGMENT}

We gratefully acknowledge the support of Saint Petersburg State University
via Grant 11.38.636.2013.

\newpage

\begin{table}[t]
\caption{Numerical values of critical exponents for $n=1$, $n=0$ and $n=-1$ found
by direct summation (DS) of the pseudo-$\epsilon$ expansions (see the text) and
of corresponding inverse series (DS$^{-1}$), by Pad\'e resummation of the series
for $\gamma$ and $\nu$, and by Pad\'e-Borel resummation of the pseudo-$\epsilon$
expansions and of their inverses using Pad\'e approximants [2/3] and [3/2]. Pad\'e
estimates presented are averaged over those given by [2/3] and [3/2] approximants.
Exact values of critical exponents and the estimates obtained from original
five-loop renormalization-group series \cite{OS2000} are also presented for
comparison.}
\label{tab1}

\begin{tabular}{|*{9}{c|}}\hline
\multicolumn{9}{|c|}{Critical exponents (CE) for various $n$.} \\ \hline
~~CE~~ & exact & ~~DS~~ & ~DS$^{-1}~ $ & ~Pad\'e~ & PB$_{[2/3]}$ &
(PB$^{-1}$)$_{[2/3]}$ & (PB$^{-1}$)$_{[3/2]}$ &~5-loop RG~\\ \hline
\multicolumn{9}{|c|}{$n=1$} \\ \hline
$\gamma$ & 1.75 & ~~1.7145~ & ~~1.7304~ & ~~1.775~~ & ~1.6105~ & ~1.7746~ &
-- & ~1.790~ \\ \hline
$\nu$    & 1 & ~~0.9067~ & ~~0.9204~ & ~0.959~ & ~0.8136~ & ~0.9652~ & -- &
~0.966~ \\ \hline
$\eta$   & 0.25 & ~~0.1372~ &   &   &   &   &   & ~0.146~ \\ \hline
\multicolumn{9}{|c|}{$n=0$} \\ \hline
$\gamma$ & ~1.34375~(43/32) & ~~1.4115~ & ~~1.4740~ & ~~1.435~~ & ~1.3804~ &
~1.4285~ & ~1.4429~ & ~1.449~ \\ \hline
$\nu$    & 0.75 & ~~0.7250~ & ~~0.7399~ & ~~0.753~~ & ~0.7069~ & ~0.7514~ &
-- & ~0.774~ \\ \hline
$\eta$   & 0.20833~(5/24) & ~~0.1204~ &   &   &   &   &   & ~0.128~ \\ \hline
\multicolumn{9}{|c|}{$n=-1$} \\ \hline
$\gamma$ & ~1.15625~(37/32)& ~~1.1917~ & ~~1.1952~ & ~~1.192~~ & ~1.1641~ &
~1.1843~ & -- & ~1.184~ \\ \hline
$\nu$    & 0.625 & ~~0.6021~ & ~~0.6183~ & ~0.606~ & ~0.5945~ & ~0.6054~ &
~0.6076~ & ~0.617~ \\ \hline
$\eta$   & 0.15 & ~~0.0793~ &   &   &   &   &   & ~0.082~ \\ \hline
\end{tabular}
\end{table}

\begin{table}[t]
\caption{Pad\'e table originating from pseudo-$\epsilon$ expansion (14) for
critical exponent $\nu$ at $n=0$. The exact value of this critical exponent
is equal to 0.75.}
\label{tab2}
\renewcommand{\tabcolsep}{0.4cm}
\begin{tabular}{|*{7}{c|}}              \hline
              L/M & 0 & 1 & 2 & 3 & 4 & 5 \\ \hline
              0 & 0.500 & 0.625 & 0.704 & 0.725 & 0.753 & 0.731 \\ \hline
              1 & 0.667 & 0.838 & 0.733 & 0.639 & 0.741 &       \\ \hline
              2 & 0.763 & 0.744 & 0.763 & 0.752 &       &       \\ \hline
              3 & 0.740 & 0.755 & 0.754 &       &       &       \\ \hline
              4 & 0.781 & 0.754 &       &       &       &       \\ \hline
              5 & 0.706 &       &       &       &       &
               \\ \hline
\end{tabular}
\end{table}

\begin{table}[t]
\caption{Pad\'e triangle for pseudo-$\epsilon$ expansion (19) of exponent
$\nu$ at $n = -1$. The exact $\nu$ value equals 0.625.}
\label{tab3}
\renewcommand{\tabcolsep}{0.4cm}
\begin{tabular}{|*{7}{c|}}
              \hline
              L/M & 0 & 1 & 2 & 3 & 4 & 5 \\ \hline
              0 & 0.500 & 0.571 & 0.606 & 0.602 & 0.614 & 0.587 \\ \hline
              1 & 0.583 & 0.640 & 0.603 & 0.605 & 0.606 &       \\ \hline
              2 & 0.619 & 0.606 & 0.610 & 0.606 &       &       \\ \hline
              3 & 0.600 & 0.609 & 0.607 &       &       &       \\ \hline
              4 & 0.618 & 0.605 &       &       &       &       \\ \hline
              5 & 0.577 &       &       &       &       &
               \\ \hline
\end{tabular}
\end{table}

\begin{table}[t]
\caption{Pad\'e table for pseudo-$\epsilon$ expansion (12) of exponent $\gamma$
at $n = 0$. The exact exponent value is 1.34375.}
\label{tab4}
\renewcommand{\tabcolsep}{0.4cm}
\begin{tabular}{|*{7}{c|}}
              \hline
              L/M & 0 & 1 & 2 & 3 & 4 & 5 \\ \hline
              0 & 1.000 & 1.25  & 1.393 & 1.412 & 1.447 & 1.383 \\ \hline
              1 & 1.333 & 1.585 & 1.414 & 1.374 & 1.424 &       \\ \hline
              2 & 1.494 & 1.439 & 1.449 & 1.429 &       &       \\ \hline
              3 & 1.414 & 1.448 & 1.441 &       &       &       \\ \hline
              4 & 1.474 & 1.430 &       &       &       &       \\ \hline
              5 & 1.326 &       &       &       &       &
               \\ \hline
\end{tabular}
\end{table}

\begin{table}[t]
\caption{Pad\'e-Borel table for pseudo-$\epsilon$ expansion of $\nu^{-1}$ at $n = 0$.
The exact exponent value equals 0.75. Some estimates are absent because corresponding
Pad\'e approximants turn out to be spoilt by "dangerous" (positive axis) poles.}
\label{tab5}
\renewcommand{\tabcolsep}{0.4cm}
\begin{tabular}{|*{7}{c|}}\hline
L/M& 0& 1 & 2 & 3 & 4 & 5\\ \hline
0 & 0.5 & 0.6058 & 0.6555 & 0.6762 & 0.6888 & 0.6954 \\ \hline
1 & 0.6667 & -- & 0.7170 & -- & 0.7145 & \\ \hline
2 & 0.7634 & 0.7449 & -- & 0.7514& & \\ \hline
3 & 0.7399 & 0.7538 & -- & & & \\ \hline
4 & 0.7814 & 0.7549 & & & & \\ \hline
5 & 0.7061 & & & & & \\ \hline
\end{tabular}
\end{table}

\begin{table}[t]
\caption{The same as Table V for $n = -1$. The exact value of $\nu$ is equal to 0.625.}
\label{tab6}
\renewcommand{\tabcolsep}{0.4cm}
\begin{tabular}{|*{7}{c|}}\hline
L/M & 0 & 1 & 2 & 3 & 4 & 5\\ \hline
0 & 0.5 & 0.5640 & 0.5903 & 0.5961 & 0.6012 & ~~~--~~~ \\ \hline
1 & 0.5833 & -- & 0.5990 & -- & 0.6009 & \\ \hline
2 & 0.6190 & 0.6072 & -- & 0.6054 & & \\ \hline
3 & 0.6000 & 0.6086 & 0.6076 & & & \\ \hline
4 & 0.6183 & 0.6059 & & & & \\ \hline
5 & 0.5766 & & & & & \\ \hline
\end{tabular}
\end{table}

\begin{table}[t]
\caption{Pad\'e-Borel triangle for pseudo-$\epsilon$ expansion of $\gamma^{-1}$ at $n = 0$.
The exact exponent value is 1.34375. Absent estimates are due to Pad\'e approximant
dangerous poles.}
\label{tab7}
\renewcommand{\tabcolsep}{0.4cm}
\begin{tabular}{|*{7}{c|}}\hline
L/M & 0 & 1 & 2 & 3 & 4 & 5 \\ \hline
0 & 1 & 1.3907 & 1.3055 & 1.3411 & 1.3622 & 1.3721 \\ \hline
1 & 1.3333 & -- & 1.3946 & -- & 1.3907 & \\ \hline
2 & 1.4942 & 1.4424 & -- & 1.4285 & & \\ \hline
3 & 1.4145 & 1.4458 & 1.4429 & & & \\ \hline
4 & 1.4740 & 1.4320 & & & & \\ \hline
5 & 1.3264 & & & & & \\ \hline
\end{tabular}
\end{table}

\begin{figure}
\begin{center}
\includegraphics[width=\linewidth]{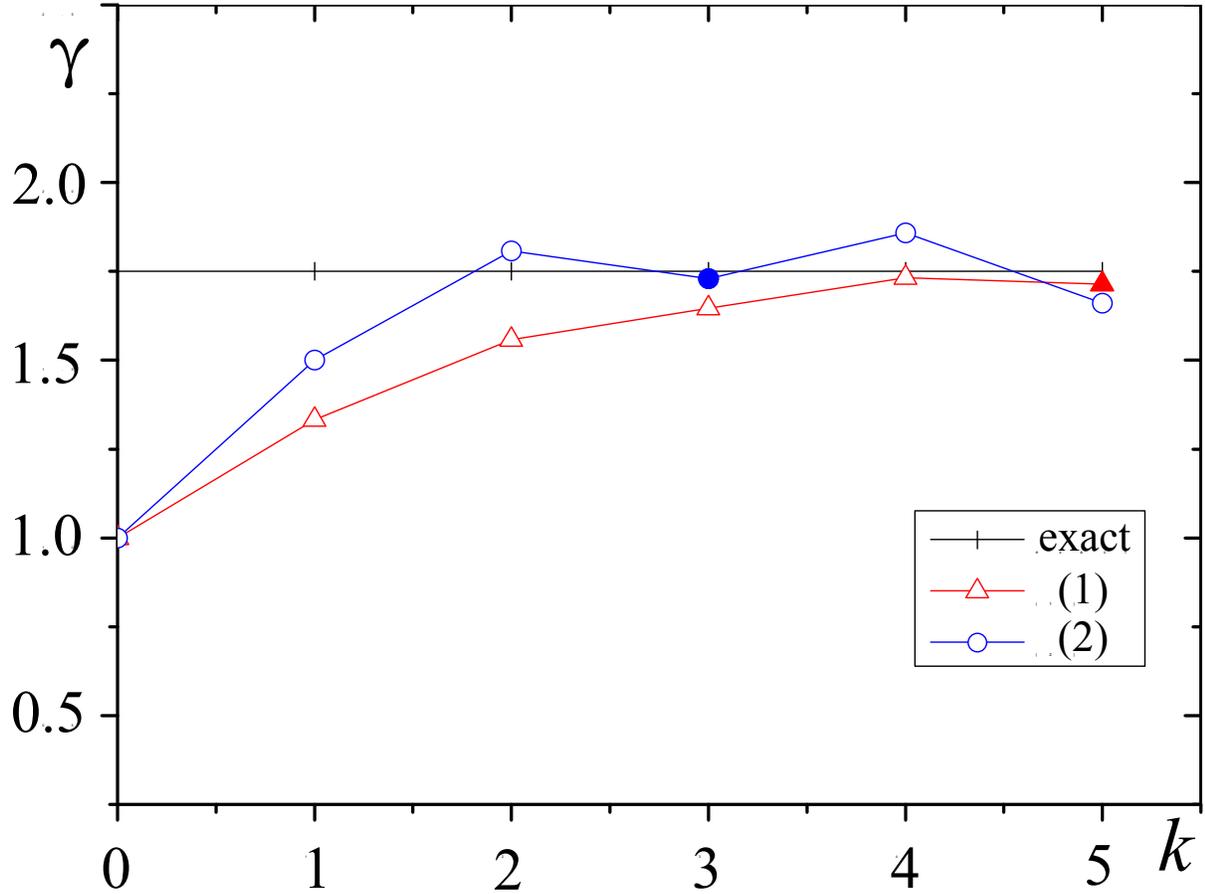}
\caption{(Color online) The values of critical exponent $\gamma$ 
for $n = 1$ as functions of the order in $\tau$ $k$ obtained by direct 
summation of pseudo-$\epsilon$ expansions (7)~(curve 1, triangles) 
and (8)~(curve 2, rounds). Horizontal line depicts the exact value.
Filled triangle and round mark the points of optimal cut off, i. e. 
the orders at which coefficients of the series finish to decrease.} 
\label{fig1}
\end{center}
\end{figure}

\begin{figure}
\begin{center}
\includegraphics[width=\linewidth]{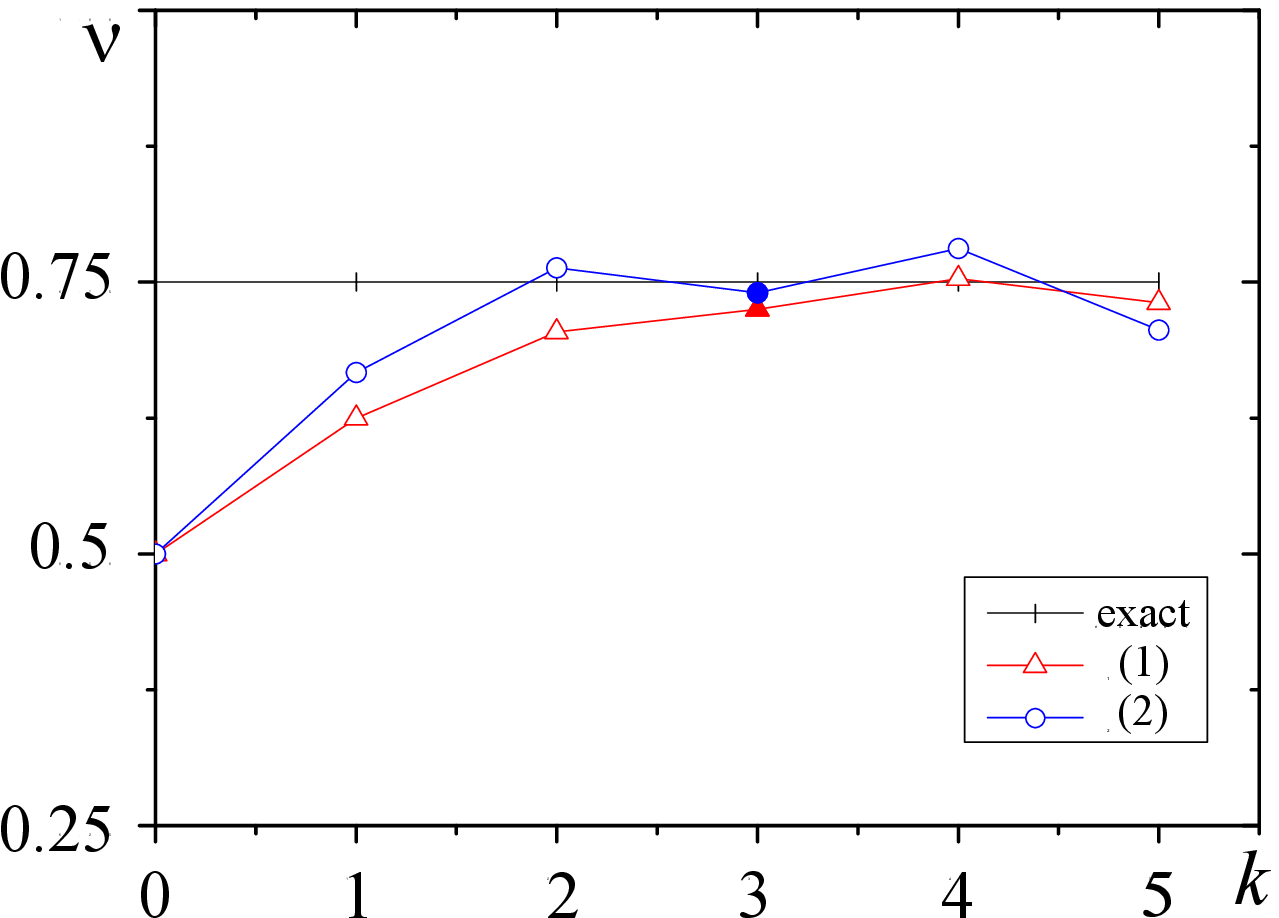}
\caption{(Color online) Same as Fig. 1, but for the exponent $\nu$ 
at $n = 0$. Triangles correspond to series (14), rounds -- to series 
(15).}
\label{fig2}
\end{center}
\end{figure}

\begin{figure}
\begin{center}
\includegraphics[width=\linewidth]{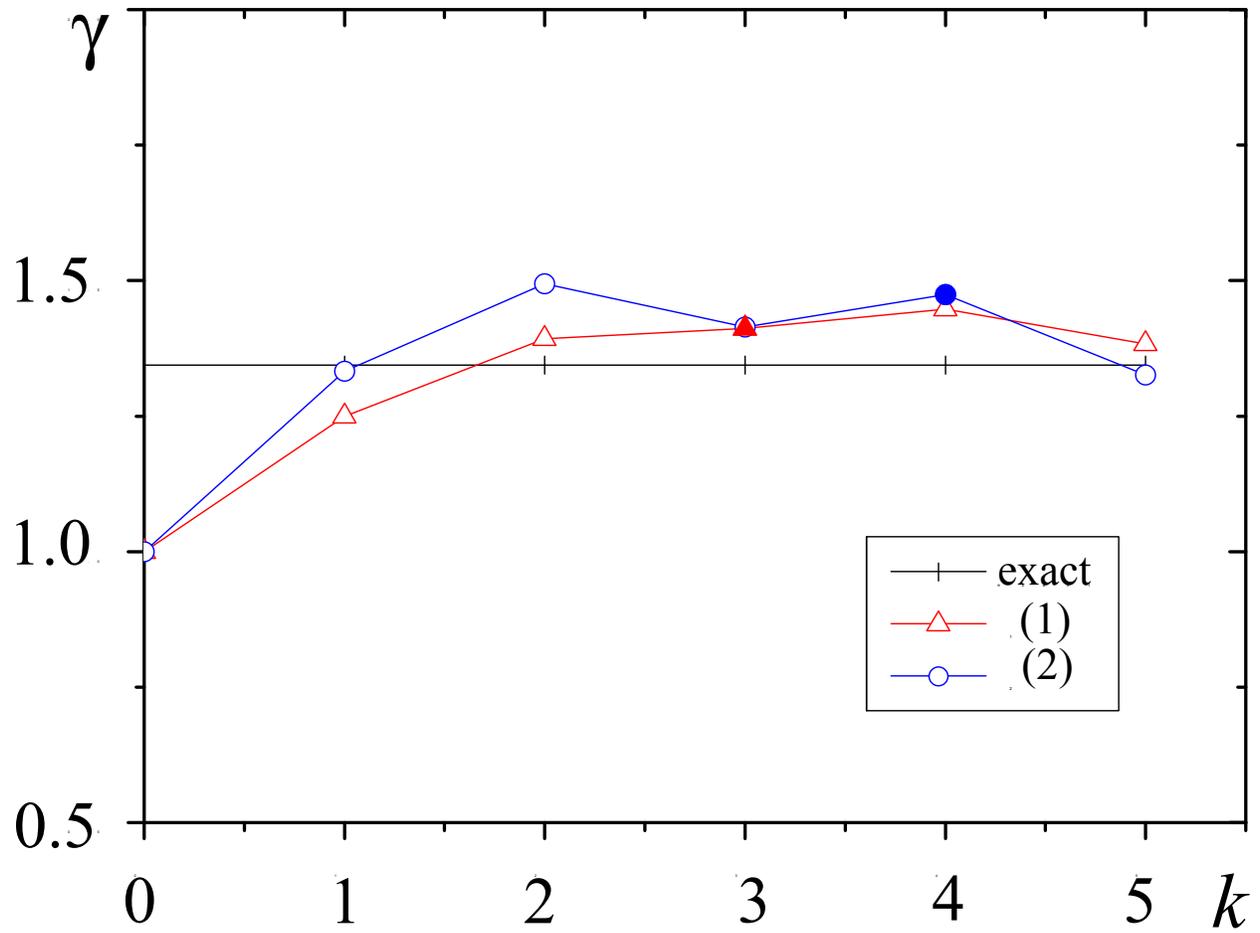}
\caption{(Color online) Critical exponent $\gamma$ at $n = 0$ as 
function of $k$ (the order in $\tau$) obtained by direct summation of 
series (12)~(triangles) and (13)~(rounds).} 
\label{fig3}
\end{center}
\end{figure}

\begin{figure}
\begin{center}
\includegraphics[width=\linewidth]{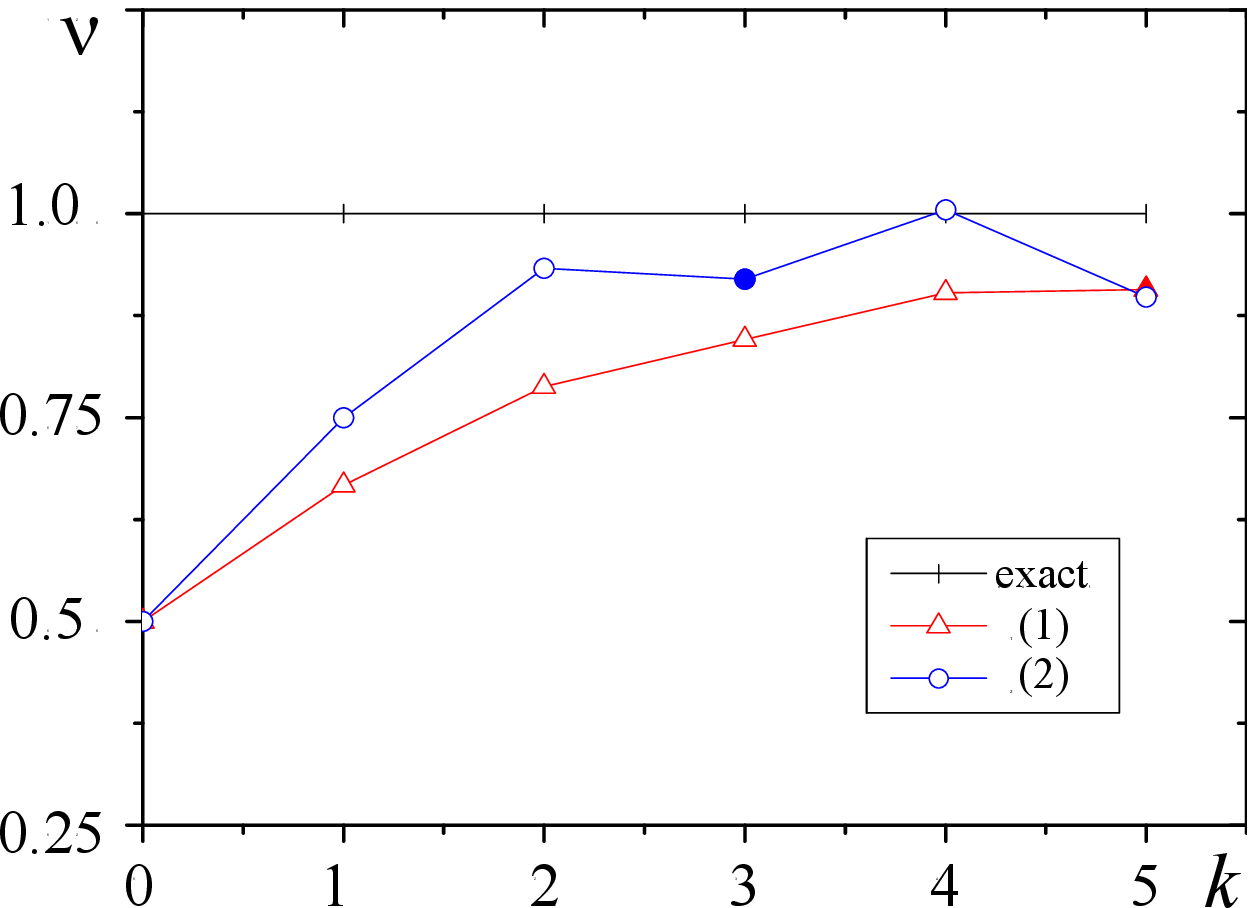}
\caption{(Color online) Same as Fig. 1, but for the exponent $\nu$ 
at $n = 1$. Triangles correspond to series (9), rounds -- to series 
(10).}
\label{fig4}
\end{center}
\end{figure}


\begin{thebibliography}{999}
\bibitem{3D} As was found very recently\cite{NS14c,NS14e,NS14h}, similar situation 
takes place for three-dimensional systems.
\end{thebibliography}
\end{document}